\documentclass[num-refs]{nbdt-article}
\usepackage{siunitx}
\usepackage{amsmath}
\usepackage{afterpage}
\usepackage{nonfloat}

\usepackage[finalnew]{trackchanges}
\addeditor{DG}
\addeditor{CB}

\papertype{Original Article}

\paperfield{Journal Section}

\title{Limitations of a proposed correction for slow drifts in decision criterion}

\abbrevs{MF: model-free}

\author[1]{Diksha Gupta}
\author[1,2]{Carlos D. Brody}

\affil[1]{Princeton Neuroscience Institute, Princeton University, Princeton, NJ 08544, U.S.A.}
\affil[2]{Howard Hughes Medical Institute, Princeton University, Princeton, NJ 08544, U.S.A.}
\corraddress{Carlos Brody, Princeton Neuroscience Institute, Princeton University, Princeton, NJ 08544, U.S.A.}

\corremail{brody@princeton.edu, dikshag@princeton.edu}
\fundinginfo{R01MH067991 from NIH and an award to CDB from the Simons Collaboration on the Global Brain}

\runningauthor{Gupta and Brody}

\begin{document}

\maketitle

\begin{abstract}
Trial history biases in decision-making tasks are thought to reflect systematic updates of decision variables, therefore their precise nature informs conclusions about underlying heuristic strategies and learning processes. However, random drifts in decision variables can corrupt this inference by mimicking the signatures of systematic updates. Hence, identifying the trial-by-trial evolution of decision variables requires methods that can robustly account for such drifts. Recent studies (Lak’20, Mendonça‘20) have made important advances in this direction, by proposing a convenient method to correct for the influence of slow drifts in decision criterion, a key decision variable. Here we apply this correction to a variety of updating scenarios, and evaluate its performance. We show that the correction fails for a wide range of commonly assumed systematic updating strategies, distorting one’s inference away from the veridical strategies towards a narrow subset. To address these limitations, we propose a model-based approach for disambiguating systematic updates from random drifts, and demonstrate its success on real and synthetic datasets. We show that this approach accurately recovers the latent trajectory of drifts in decision criterion as well as the generative systematic updates from simulated data. Our results offer recommendations for methods to account for the interactions between history biases and slow drifts, and highlight the advantages of incorporating assumptions about the generative process directly into models of decision-making.
\end{abstract}

\section{Introduction}
Animals’ choices in perceptual decision-making tasks often display a dependence on the recent history of stimuli, choices, and outcomes. This dependence is thought to arise from systematic updating of decision variables from trial to trial. These updates may reflect ongoing learning \cite{dayan_decision_2008}, for instance an agent learning to perform a perceptual categorization task might update its beliefs about the prior probabilities of the different categories \cite{yu_dynamics_2009, zhang_sequential_2014}, the category boundary separating them \cite{drugowitsch_learning_2019, mendonca_impact_2020}, or the values of the available actions \cite{lak_dopaminergic_2020, lak_reinforcement_2020, pisupati_lapses_2021}. Alternatively, systematic updates may reflect heuristic strategies adopted by decision-makers due to resource constraints or mismatched objectives \cite{gigerenzer_heuristic_2011, abrahamyan_adaptable_2016, gardner_optimality_2019}. For example, a reward-seeking agent might be prone to repeating rewarded actions and avoid unrewarded ones, i.e. follow a win-stay lose-switch strategy, similarly high costs of motor switching may encourage an agent to repeat previously chosen actions, yielding stay biases. These learning processes and heuristic strategies are often identified by the distinct patterns of choices they predict.

In addition to these systematic updates, decision variables may drift randomly from trial to trial \cite{dutilh_how_2012, lak_reinforcement_2020, ashwood_mice_2022, cowley_slow_2020, hennig_learning_2021, roy_extracting_2021, lyamzin_probabilistic_2021}. These “unsystematic” drifts may arise from history-independent fluctuations in internal states such as attention, arousal and motivation, or from other unknown sources of noise \cite{renart_variability_2014}. However, despite their history independence, unsystematic drifts may nevertheless produce correlations in choices that obscure the effect of systematic updates and thereby complicate their inference \cite{lak_reinforcement_2020, sugawara_dissociation_2021}.

An important recent study \cite{lak_reinforcement_2020} considered the challenge posed by unsystematic drifts in one key decision variable, the decision criterion, i.e. the threshold for choosing one alternative over the other. The authors showed that a slow drift in criterion produces an apparent history dependence that mimics the signatures of updates from a systematic learning process. Hence, in order to remove the influence of these slow drifts, they proposed a model-free (MF) correction \cite{lak_reinforcement_2020}, similar in spirit to an approach employed by \cite{mendonca_impact_2020}. They reasoned that slow drifts would produce correlations in choices across multiple successive trials. Thus behavior in a given trial would be correlated with both immediately previous and immediately future trials, giving the appearance that experience in the current trial would not just influence future choices (a causal process), but would also spuriously influence past choices (an acausal, therefore not biologically plausible process). Hence, they posited that the effect of slow drifts of decision criterion can be removed by \textit{subtracting the influence of the current trial on past choice (the acausal component) from the influence of the current trial on future choice}. The simplicity of this correction is appealing and has already invited other authors to use it on their datasets. Indeed, recently the International Brain Laboratory reported that their dataset displayed win-stay lose-stay behavior, but when they accounted for possible slow drifts using the MF correction, they revealed instead a win-stay lose-switch strategy \cite{the_international_brain_laboratory_standardized_2021}.

While \cite{lak_reinforcement_2020} demonstrated convincingly that the MF correction can remove the influence of random slow drifts, they did so in the absence of any systematic updates of decision-variables. It remains to be shown if the MF correction is robust to the presence of learning or other heuristic strategies. If the MF correction indiscriminately removes correlations across choices, including those expected from systematic trial-by-trial updating of decision-relevant variables, that would considerably undermine its usefulness. 

Here, using simulations we show that the proposed model-free (MF) correction fails in the presence of systematic trial-by-trial updating, yielding potentially misleading conclusions about the nature of underlying behavioral strategies. We show that the correction erroneously removes the effect of trial-by-trial updates that produce correlations across choices, thereby misidentifying a variety of update strategies as belonging to a narrow subspace of win-stay lose-switch or win-switch lose-stay biases. We advocate for an alternate, model-based approach for disambiguating systematic updates from random drifts, and demonstrate the success of this approach by fitting synthetic data. Finally, we apply the two approaches to real data from rodents \cite{brunton_rats_2013} and demonstrate that whereas the model-based approach proposed here preserves individual variability in systematic history biases across animals (while removing the influence of slow drifts), the previous model-free approach collapses this variability into a small subset of apparent strategies. Our results highlight the importance of evaluating assumptions underlying model-free corrections, and offer arguments for incorporating assumptions about the generative process directly into model fitting.

\section{Results}
\subsection{Correction in action}

\afterpage{
\refstepcounter{figure}
\begin{center}
\includegraphics[width=11cm]{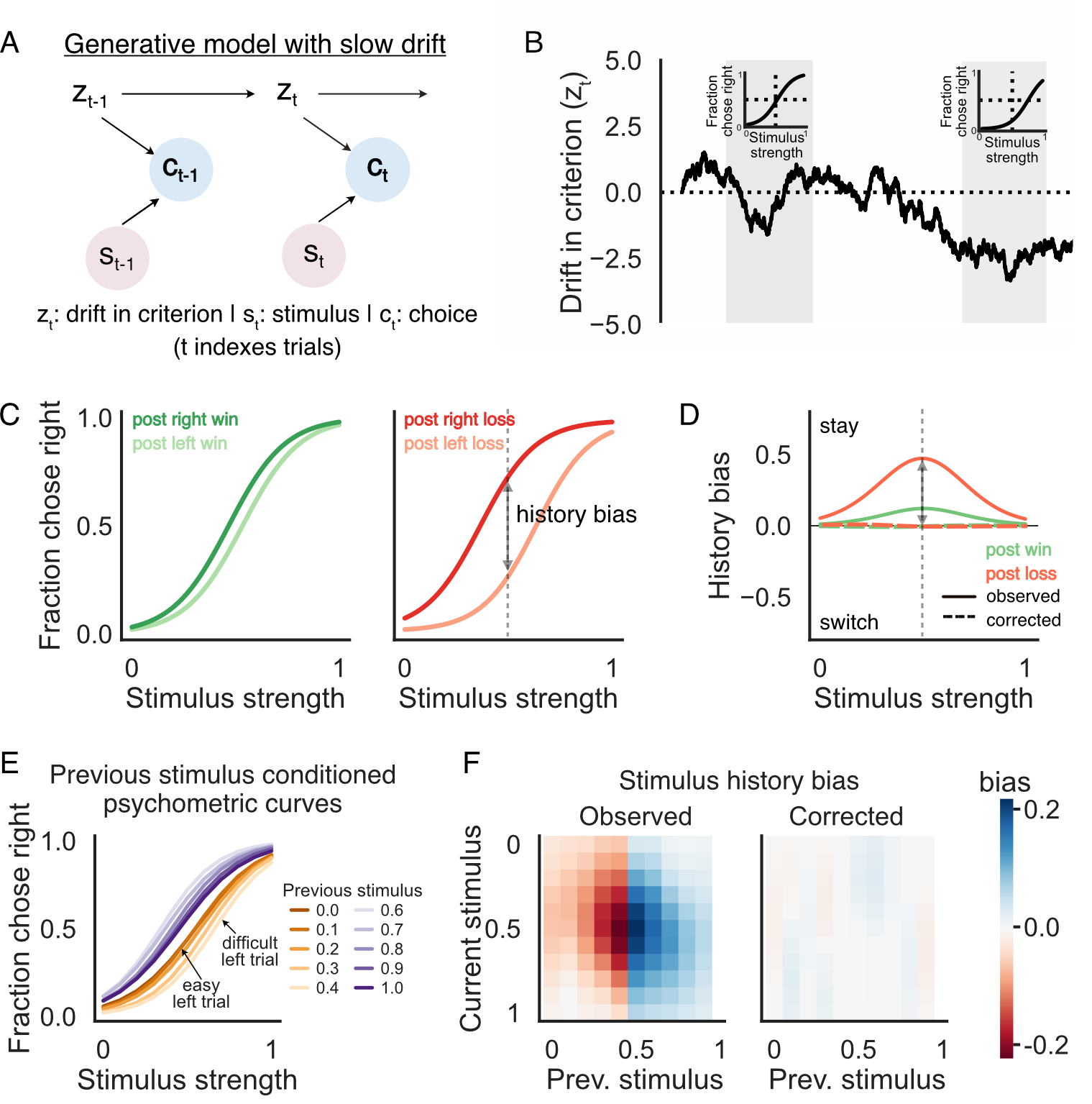}
\end{center}
\figcaption*{
\justifying
\textbf{FIGURE 1 : Correction for removing influences of slow drift in psychometric parameters on choice data (as proposed in \cite{lak_reinforcement_2020})}
\small{
\textbf{A.} Generative model of choices based on signal detection theory: Each trial’s choice $c$ is made by comparing a noisy sample of that trial’s stimulus $s$ to a slowly drifting criterion (drift in the criterion represented by $z$, trials indexed by $t$) 
\textbf{B.} Trajectory of the slow drift in criterion over a period of 5000 trials revealing extended periods of consistent choice bias. Insets: psychometric curves for the trials highlighted in grey when choices are on average unbiased (left) and biased (right)
\textbf{C.} Psychometric curves (fraction of rightward choices as a function of the current trial’s stimulus strength) conditioned on the previous trial’s choice and outcome, displaying an apparent win-stay lose-stay bias. Colors indicate previous outcome (green - win, red - loss) and lightness indicates previous choice (dark - right, light - left)
\textbf{D.} History bias, measured as the difference between conditioned psychometric curves following right and left wins (green) or losses (red). Solid lines indicate observed history bias, displaying an apparent small win-stay, larger lose-stay effect. Dashed lines indicate history bias after applying the MF correction, which removes the spurious choice correlation created by slow drift.
\textbf{E.} Psychometric curves conditioned on the previous trial’s stimulus when the previous trial was rewarded. Curves display an apparent dependence on past stimulus strength, and have bigger shifts following hard trials compared to easy trials. Colors indicate previous stimulus category (violet - rightward, amber - leftward) and lightness indicates strength of the stimulus (dark - high stimulus strength or easy trials, light - low stimulus strength or hard trials)
\textbf{F.} (Left) Stimulus history bias, measured as the difference between psychometric curves conditioned on previous stimulus and average psychometric curve. Biases are larger following hard trials (color gradient along x-axis) and disproportionately affect the performance on hard trials (color gradient along y-axis). Color indicates the direction of induced bias (blue - towards right, red - towards left) (Right) Stimulus history biases after applying the model-free (MF) correction, which removes the spurious dependence due to slow drifts in decision criterion.
}}
\label{fig1}
\vspace{1cm}
}

We begin by reproducing the results from \cite{lak_reinforcement_2020} that the proposed model-free (MF) correction removes spurious choice correlations that are introduced by slow drifts in decision criterion. We consider a generative model \add[DG]{(}\hyperref[fig1]{Figure 1A}) in which the stimulus affects the choice of an agent in accordance with signal detection theory (see Methods \remove[DG]{, Figure 1A}). On any given trial $t$, the agent compares a noisy perceptual sample, centered around the true stimulus $s_t$ with logistic noise (scale $\beta_S$) to a decision criterion $b_t$, and makes a rightward choice if the sample exceeds the criterion and leftward choice otherwise. \change[DG]{Therefore,}{The} probability of making a rightward choice is \add[DG]{therefore} given by the following logistic function: 
$$ p(c_t = 1| s_t,b_t) = \frac{1}{1+e^{-(s_t - b_t)/\beta_S}}$$
The decision criterion $b_t$ of this observer varies from trial-to-trial relative to a fixed baseline $\beta_0$ due to unsystematic variations arising from random drift $z_t$:
$$ b_t = \beta_0 + z_t$$
This drift $z$ in decision criterion $b$ (\hyperref[fig1]{Figure 1B}) was simulated according to the following autoregressive process (discrete time Ornstein-Uhlenbeck process):
$$z_t = (1-\lambda)z_{t-1} + \sigma_d\epsilon_t$$
where $\epsilon_t$ is an i.i.d sequence of standard Gaussian random variables and $\sigma_d$ sets the standard deviation of the Gaussian noise. The decay rate $\lambda$ was fixed to a small value ($\lambda = 5e-4$) to slightly dampen the pure Brownian motion thereby reducing the odds of the criterion drifting too far from the true boundary.

We assume that there is no learning or systematic trial-by-trial updating of any decision-related variables. In spite of that, the slowly drifting criterion produces correlations in the agent’s choices across trials, such that when the psychometric curve is conditioned on previous trial’s outcome and choice, it appears that the agent is following a small win-stay larger lose-stay strategy (\hyperref[fig1]{Figure 1C})-- in other words, it appears that subjects use the experience of a trial’s outcome to affect future trials, when in fact reward outcomes do not affect the decision process at all. The psychometric curves following both correct and incorrect rightward choices are biased towards rightward choices and vice versa following leftward choices. We summarize these apparent history effects by taking the difference between conditioned psychometric curves following right and left wins/losses. Therefore, following correct trials the history bias is given according to:
$$\text{Bias}^{PC}_{t|t-1} = \frac{1}{2} \bigl(p(c_t = 1|s_t, o_{t-1} = 1, c_{t-1} = 1) - p(c_t = 1|s_t, o_{t-1} = 1, c_{t-1} = 0)\bigr) $$ 
where $o_{t-1}=1$ if the trial $t-1$ is a win and is $0$ otherwise (PC refers to post-correct). And to compute the bias following error trials (PE) we instead condition on trial $t-1$ being an incorrect trial or $o_{t-1} = 0$:
 $$\text{Bias}_{t|t-1}^{PE} = \frac{1}{2}\bigl(p(c_t = 1|s_t, o_{t-1} = 0, c_{t-1} = 1) - p(c_t = 1|s_t, o_{t-1} = 0, c_{t-1} = 0)\bigr) $$
This metric is positive if the behavior has a stay bias and is negative when there is a tendency to switch. Even though the drift in criterion is independent of the previous trial’s outcome, the stay bias appears previous outcome-dependent - its magnitude is smaller following correct choices than erroneous choices (\hyperref[fig1]{Figure 1D}, solid lines). This effect arises because the agent is more prone to errors and more likely to have a persistent choice bias when the criterion has drifted further away from the true boundary. 

Further, we examine how recent sensory history affects future choices under such a generative model and observe that again, the slow drifts in criterion produce the semblance of systematic trial-by-trial updating (\hyperref[fig1]{Figure 1E,F}; replication of Figure 2, figure supplement 1 from \cite{lak_reinforcement_2020}). It appears as though the agent's choice on the current trial is modulated by the previous trial’s stimulus difficulty (\hyperref[fig1]{Figure 1E}). Following correct trials the agent has a higher propensity to repeat the correct response to the previous trial’s category if the previous trial was difficult. In contrast if the past trial was an error the bias is higher following an easy trial (not shown). \add[DG]{This apparent difficulty-dependent stimulus history bias (resembling confidence-guided updating) arises because of the statistics of occurrence of correct trials in the presence of drifts in criterion. Easy trials are resilient to drifts in criterion and are just as likely to be correct both when the drifts favor the correct option and when they are biasing the decision against it, whereas hard trials have a higher tendency of being correct when the drift in criterion is biased towards the correct option. As a result, trials following a correct hard trial are sampled from a distribution in which the median value of the criterion is biased, hence the conditioned psychometric curve has a large bias. Whereas, trials following an easy trial results from choices with relatively unbiased criterion values, hence the conditioned psychometric curve has a smaller bias.}

We summarize these apparent stimulus history biases following rewarded trials for each pair of previous and current stimuli in \hyperref[fig1]{Figure 1F} (left panel). \change[DG]{The bias is }{These biases are} computed by subtracting the psychometric curve computed \change[DG]{from}{using} all trials from the psychometric curves conditioned on the previous trial’s stimulus strength (similar to \cite{lak_reinforcement_2020}):
$$\text{Bias}_{s_t|s_{t-1}} = \frac{1}{2} \bigl(p(c_t = 1|s_t,s_{t-1}, o_{t-1} = 1) - p(c_t = 1|s_t)\bigr) $$ 

Next, we apply the MF correction to this synthetic data. The MF correction exploits the idea that the effect of slow drifts should be similar for at least a small stretch of trials $(n>=2)$ and any influence of future rewards on past choices is acausal, likely due to slow drifts and needs to be removed. \add[DG]{Mendonca et al. (2020)} \cite{mendonca_impact_2020} proposed the variant of the correction for choice-outcome biases and \add[DG]{Lak et al. (2020)} \cite{lak_reinforcement_2020} for stimulus history biases. We describe and test the efficacy of both these variants here. For post-correct (PC) history bias, the MF correction entails subtracting the (spurious) bias due to trial $t-1$ on the past trial $t-2$ from the bias due to trial $t-1$ on the future trial $t$:
$$ \text{Corrected Bias}^{PC}_{t|t-1} = \text{Bias}^{PC}_{t|t-1} - \text{Bias}^{PC}_{t-2|t-1}$$
Post-error history bias is corrected for slow drifts in a similar manner. The MF correction for stimulus history effects stipulates that the effect of slow drifts in decision criterion can be removed by performing an equivalent operation on inferred stimulus history bias:
$$ \text{Corrected Bias}_{s_t|s_{t-1}} = \text{Bias}_{s_t|s_{t-1}} - \text{Bias}_{s_{t-2}|s_{t-1}}$$
When applied to synthetic data, the MF correction removes these apparent choice history effects caused by the drifting criterion and successfully recovers the systematic component of past choice’s influence on subsequent choice i.e. no influence (\hyperref[fig1]{Figure 1D}, dashed lines). Similarly, specious stimulus history effects could be successfully removed by applying the MF correction (\hyperref[fig1]{Figure 1F}, right panel).

\subsection{Correction in misaction}

\afterpage{
\refstepcounter{figure}
\begin{center}
\includegraphics[width=12.5cm]{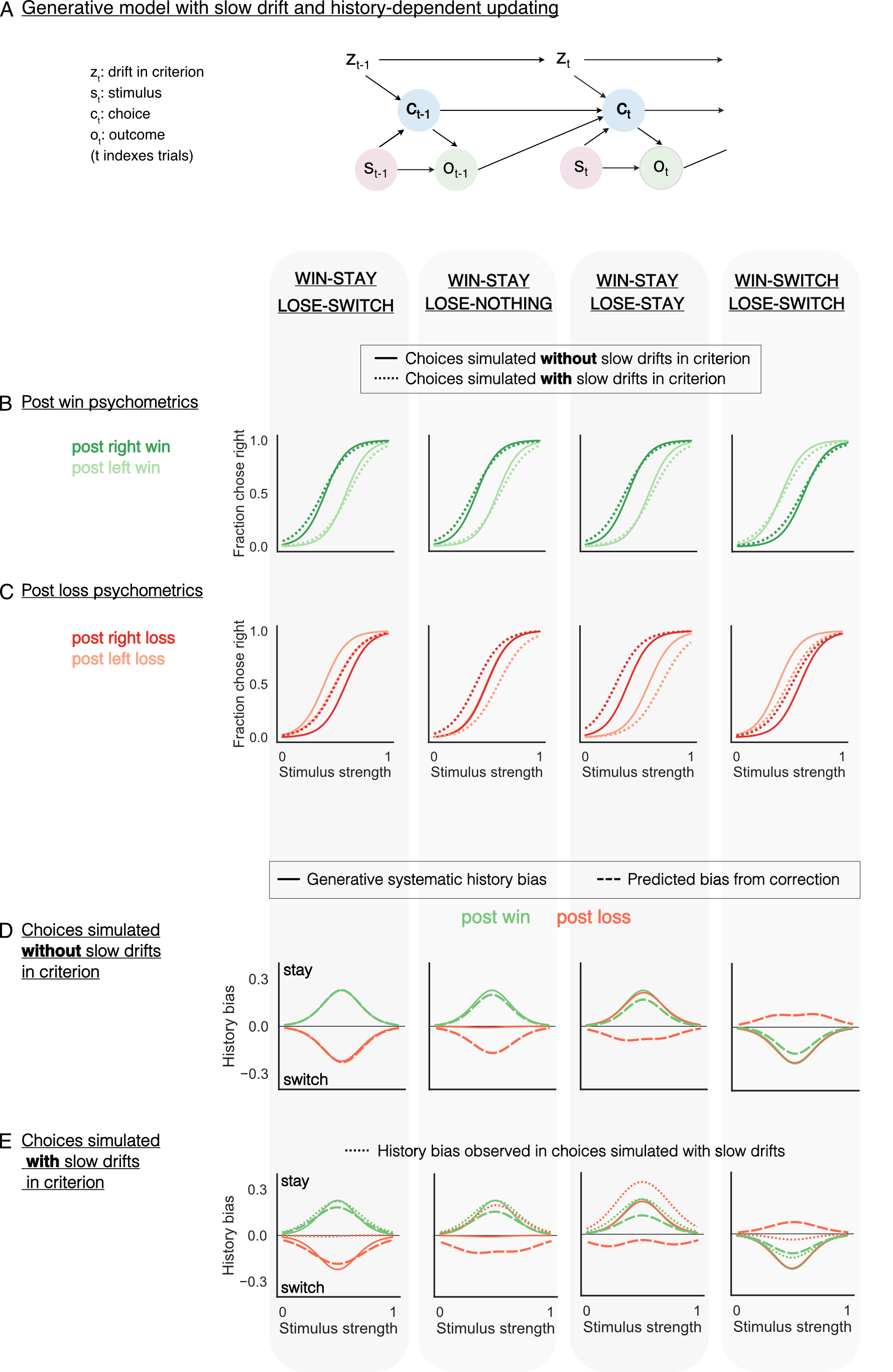}
\end{center}
\figcaption*{
\justifying
\textbf{FIGURE 2 : Correction fails to recover generative parameters when decision variables are actively updated from trial-to-trial}
\small{
\textbf{A.} Generative model in which the current trial’s choice $c_t$ is influenced by the previous choice $c_{t-1}$ and previous outcome $o_{t-1}$, in addition to the current trial’s stimulus $s_t$ and a slow drift in criterion $z_t$ . In some scenarios, the previous outcome does not influence the choice on the next trial e.g. win-stay lose-stay or win-switch lose-switch. 
\textbf{B.} Psychometric curves post rewarded trials, conditioned on previous trial’s choice for a model with win-stay, lose-switch (column 1), win-stay lose-nothing (column 2), win-stay lose-stay (column 3) and win-switch lose-switch (column 4) biases. Solid lines represent psychometric curves without slow drift and dotted lines with slow drift. Addition of slow drift gives rise to worse perceptual sensitivity.
\textbf{C.} Same as B but for trials following losses or unrewarded trials. Addition of slow drift masks lose-switch effects and overlays a stay bias across all updating strategies.
\textbf{D.} History biases following wins (green) and losses (red) without slow drifts in criterion. Solid: generative systematic history biases, dashed: estimates from MF correction. MF correction nearly recovers the true generative post win history biases (dashed vs solid green line) across different updating strategies. For post loss biases (dashed vs solid red line), the MF correction has no effect for win-stay lose-switch bias (column 1), but produces spurious lose-switch effects for win-stay lose-nothing (column 2) and win-stay lose-stay (column 3) biases. For win-switch lose-switch (column 4) bias the correction returns a lose-stay bias. 
\textbf{E.} Same as D but for history biases following wins (green) and losses (red) with slow drifts in criterion. The correction fails to recover the generative history biases (dashed vs solid lines) in this case as well. \add[DG]{Dotted lines depict the history bias observed in choices simulated with drift (corresponding to dotted lines in 2B,C).}
}}
\label{fig2}
\vspace{1cm}
}

In the past section, the generative model lacked active trial-by-trial updating and hence the correlations in choice were entirely due to the slow drift in decision criterion. Next we consider a more complex scenario, one in which the decision variables are systematically updated from trial-to-trial due to ongoing learning or other heuristic strategies in addition to the random drifts in decision criterion, and we determine how the MF correction performs under these circumstances. 

Previous studies have shown that a large subset of possible learning strategies can be approximated by logistic regression models that directly represent the influence of reward and choice history on future choices \cite{katahira_relation_2015, miller_predictive_2019}. For this reason, in this study we explore the space of possible choice and outcome history effects instead of examining the correction’s effect in the presence of individual learning algorithms. We parametrically vary the bias induced by past choice ($c$) and outcome ($o$) and study the effect of MF correction both in the presence and absence of slow drifts ($z$) of decision criterion (\hyperref[fig2]{Figure 2A}). We find that while the MF correction can correctly recover a symmetric win-stay lose-switch history dependence, it fails to recover the generative history-dependence in other scenarios.

In order to simulate choices from this family of generative models, we update the randomly drifting decision criterion $b_t$ every trial based on the recently observed choice and outcome from the previous trial. Therefore, on any trial the decision criterion is given by:
$$ b_t = z_t + \beta_0 + \beta_C(\mathbb{I}_{t-1}^{RC} - \mathbb{I}_{t-1}^{LC}) + \beta_E(\mathbb{I}_{t-1}^{RE}-\mathbb{I}_{t-1}^{LE})$$
where $\mathbb{I}^{RC}_{t-1}$, $\mathbb{I}^{LC}_{t-1}$, $\mathbb{I}^{RE}_{t-1}$ and $\mathbb{I}^{LE}_{t-1}$ are indicator variables denoting successes or failures on the previous trial ($t-1$)  following rightward and leftward choices (RC refers to right correct, LC left correct, RE right error and LE left error). $\beta_C$ and $\beta_E$ determine how much previous trial’s outcome modulates the criterion ($\beta_C$ following correct trials and $\beta_E$ following error trials). As before, $\beta_0$ is the fixed baseline and $z_t$ denotes unsystematic variations arising from random drift.

First, we simulate data with a symmetric win-stay lose-switch rule, such that positive outcomes or correct trials bias the agent towards repeating the choice that led to them, and negative outcomes do the opposite i.e. $\beta_C = 1$, $\beta_E = -1$ and $z_t = 0$  $\forall$ $t$ (\hyperref[fig2]{Figure 2B,C}, first column, solid lines). When, in addition to win-stay lose-switch, the decision criterion slowly drifts over trials ($z_t \neq 0$), the systematic lose-switch rule is obscured and the choice behavior seems to have no updating following error trials (\hyperref[fig2]{Figure 2B,C} first column, dotted lines). Moreover, the slope of the psychometric curve (also referred to as perceptual sensitivity) appears shallower. 

We apply the MF correction to this data and examine the observed and corrected history biases with and without an underlying slow random drift in the decision criterion of the generative process. In the absence of slow drifts, as expected, the MF correction (\change[DG]{dotted}{dashed} lines) does not change our estimate of history biases and we recover the generative win-stay lose-switch bias (\hyperref[fig2]{Figure 2D}, first column). But even in the presence of slow drifts, the MF correction successfully nullifies its contaminating influence on history bias and we approximately recover the generative win-stay lose-switch bias \hyperref[fig2]{Figure 2E} first column). 

Next, we consider a scenario in which the magnitude of biases following wins (i.e.\remove[DG]{.} rewarded trials) versus losses (unrewarded trials) are asymmetric. We eliminate the bias induced by negative outcomes and study win-stay lose-nothing ($\beta_C = 1, \beta_E = 0$; \hyperref[fig2]{Figure 2B,C} second column). Here the slow drift masks the true post-error biases and instead accentuates stay biases (\hyperref[fig2]{Figure 2B,C dotted lines}). In the absence of slow drift, we can directly observe the generative bias, therefore in principle the MF correction should not affect our estimate of history biases. But instead we observe that the MF correction distorts our estimate of history biases, skewing them \change[DG]{towards win-stay lose-switch like updating}{to resemble a win-stay lose-switch updating strategy} (\hyperref[fig2]{Figure 2D} second column). 

\add[DG]{This distortion arises from the interaction of average choice statistics cast by the win-stay lose-nothing strategy with the functional form of the correction. The MF correction dictates that we should be able to recover the systematic influence of a trial on the next trial by subtracting the acausal influence (influence of a trial on the previous trial) from the observed influence. In this case, the observed influence on error trials is zero (lose-nothing). The acausal influence however is non-zero despite the absence of slow drifts - this is because on average there are more correct trials than error, and hence more often than not the current choice is a repetition of past choice (due to win-stay), therefore the acausual influence is inferred to be a stay bias. So when we apply the correction and subtract this stay bias from zero we recover a spurious lose-switch bias on error trials.} In the presence of slow drift too, the MF correction is unable to recover the generative systematic history bias (\hyperref[fig2]{Figure 2E}, second column solid vs dashed) and instead inaccurately estimates a stay bias following wins and a switch bias following losses.

When there is no dependence on previous outcomes, and the bias following both correct and error trials are towards the previous choice i.e. a win-stay lose-stay bias ($\beta_C = 1, \beta_E = 1$; \hyperref[fig2]{Figure 2B,C} third column), slow drifts in decision criterion exacerbate the stay effects following error trials. In synthetic data generated without any drifts in decision criterion but with stay bias in effect, the MF correction spuriously infers a win-stay lose-switch strategy (\hyperref[fig2]{Figure 2D} third column). A similar failure in recovering the generative parameters is observed in the presence of slow drifts (\hyperref[fig2]{Figure 2E} third column). 

Next we examine the effect of the MF correction when the data is generated from an outcome independent switch rule i.e. win-switch lose-switch ($\beta_C = -1, \beta_E = -1$; \hyperref[fig2]{Figure 2B,C} last column). In this case too, slow drift obscures the true post-error effects and overlays a stay bias on it. Again, we find that the MF correction fails to recover generative biases both in the presence and absence of slow drifts - it instead returns a win-switch lose-stay like bias (Figure \hyperref[fig2]{Figure 2D,E} last column).

\subsection{A model-based solution}

\afterpage{
\refstepcounter{figure}
\begin{center}
\includegraphics[width=12cm]{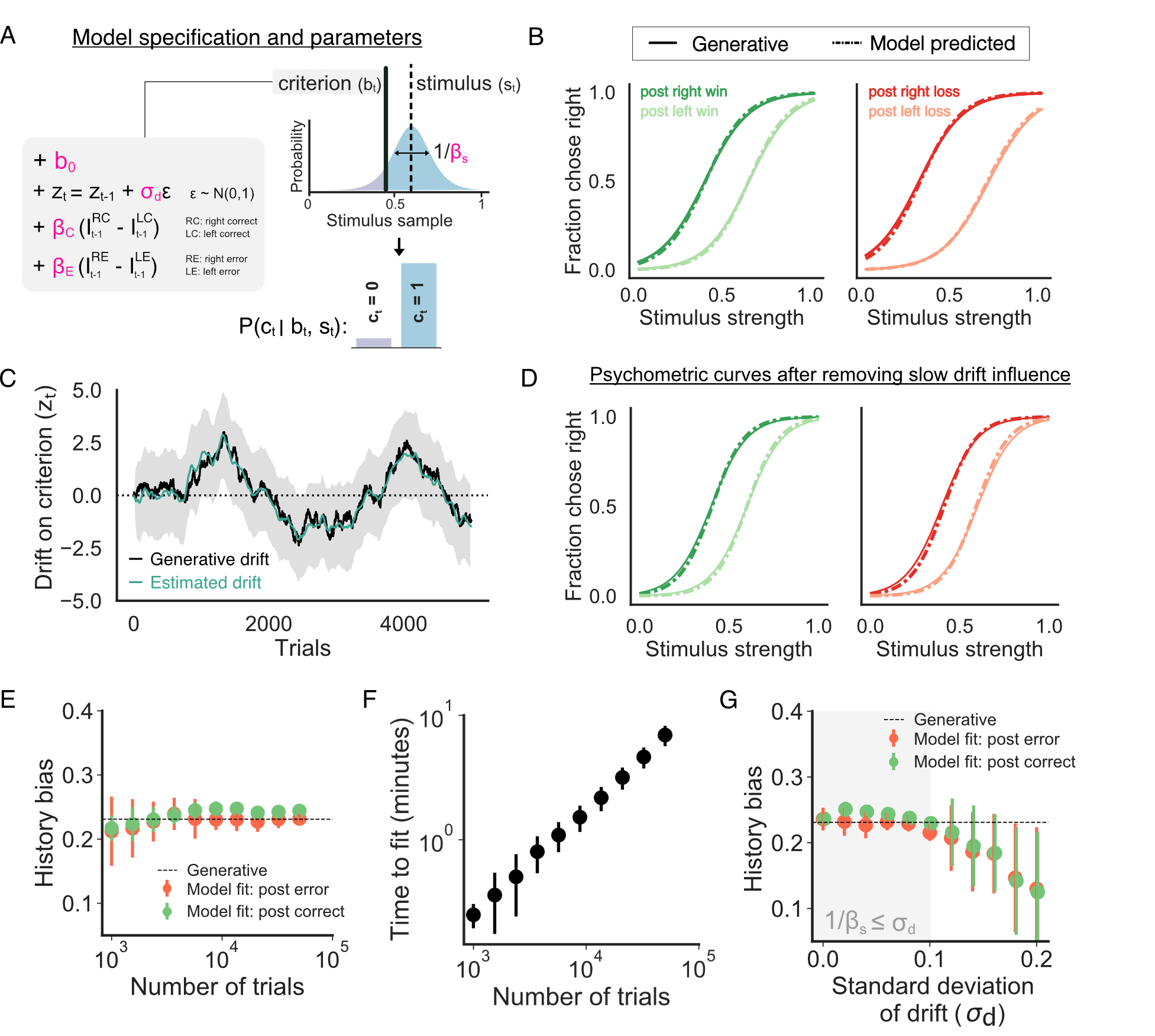}
\end{center}
\figcaption*{
\justifying
\textbf{FIGURE 3 : Fitting the model helps disambiguate systematic history biases from effects of slow drift}
\small{
\textbf{A.} Schematic of the model with parameters highlighted in pink. (Top right) Logistic distribution of noisy samples perceived by a subject for a given true stimulus $s_t$ (dotted line). The probability of the current trial’s choice $c_t$ being rightward/leftward  (shaded blue, bottom right) is given by the mass to the right/left of the criterion $b_t$ (solid black line), which evolves according to the equations shown in the box (left - $z_t$ is drift in criterion, $\mathbb{I}_{t-1}$ is an indicator variable for the previous trial’s choice, outcome). 
\textbf{B.} Psychometric curves from simulated data (solid lines) shown alongside fits (dashed-dotted lines) to a model containing both history biases and slow drifts in criterion
\textbf{C.} Inferred slow drifts from model fits (grey) compared to the true generative slow drift (black) showing good correspondence\add[DG]{Shaded error bars indicate two standard deviations of the posterior.}
\textbf{D.} Inferred history biases from model fits (dashed-dotted lines) compared to the true history biases (solid lines) showing good correspondence. Note: these are hypothetical psychometric curves that would have been observed in the absence of slow drift, and hence not directly observable from the data as in B)
\add[DG]{
\textbf{E.} Mean inferred systematic history bias (at the point of indifference) following correct (green) and error (red) trials as a function of the number of trials (in log-scale). Dotted line represents the  generative systematic bias (symmetric win-stay lose-stay). Error bars denote standard deviation across 20 independent simulated datasets.
\textbf{F.} Time taken to fit the model as a function of the umber of trials (both axes in log-scale). Error bars denote standard deviation across 20 independent simulated datasets.
\textbf{G.} Mean inferred systematic history bias (at the point of indifference) following correct (green) and error (red) trials as a function of standard deviation of the drift. Dotted line represents the generative systematic bias (symmetric win-stay lose-stay). Error bars denote standard deviation across 20 independent simulated datasets.}
}}
\label{fig3}
\vspace{1cm}
}

The foregoing analysis demonstrates that the proposed MF correction produces inaccurate results when the systematic trial-to-trial updating of history bias deviates from a symmetric win-stay lose-switch strategy. The correction assumes that correlations in choice arise from processes unrelated to deterministic trial-by-trial updating - an assumption that is untrue for many learning or heuristic algorithms. This could be remedied by explicitly estimating the contributions of both slow drifts and systematic updates to choice behavior.  

Here we describe one such approach for the simulations discussed in the previous section. We infer the parameters governing behavior by fitting the choices to the generative model at hand: a signal detection theory observer with logistic noise, systematic trial-history dependent biases and drifts in criterion (\hyperref[fig3]{Figure 3A}, see Methods). We fit the parameters $\Theta = \{\sigma_d, b_0, \beta_S, \beta_C, \beta_E\}$ of the model to choices using the Expectation Maximization algorithm with a Laplace approximation of the posterior over the latent state (drifting variable; \cite{macke_empirical_2011} as implemented in the ssm library \cite{linderman_ssm_2020}. 

We simulate data from an agent that follows a win-stay lose-stay strategy and has random drifts in its decision criterion (similar to \hyperref[fig2]{Figure 2}, third column). The MF correction is ineffective in recovering the generative parameters in this case. The model successfully captures trends in the data (\hyperref[fig3]{Figure 3B}: Fits (dashed-dotted lines) plotted alongside observed choice and outcome conditioned psychometric curves from the simulated data (solid lines)). We next examine if the model is able to tease apart the relative influences of slow drift and trial history bias on choice behavior and indeed, the inferred most likely trajectory of the latent drifting variable (\hyperref[fig3]{Figure 3C}, grey line) closely tracks the true simulated drift (\hyperref[fig3]{Figure 3C}, black line). Furthermore, the model correctly infers the true parameters governing the systematic component of the behavior. We demonstrate this by simulating choices in the absence of slow drifts with both the generative parameters and estimated parameters. The previous choice and outcome conditioned psychometrics from the generative and estimated parameters show good correspondence (\hyperref[fig3]{Figure 3D}). Additionally, this method helps recover the true perceptual sensitivity, which is otherwise confounded by the drifting biases (as in \hyperref[fig2]{Figure 2B,C}).

\add[DG]{Further, we explore the performance of this model-based approach in low trial-count and high drift-noise regimes. We simulated datasets with systematic win-stay lose-stay biases (same as Figure 3B), with varying trial counts ranging from 1000 to 50000. We then fit these datasets with the model, and examined the estimated systematic component of history bias (measured at the point of indifference, i.e. the stimulus strength at which the probability of making rightward and leftward choices is equal) as well as the time to fit. Even for trial counts as low as 1000, the model accurately infers the win-stay lose-stay bias with reasonable precision (Figure 3E), with fitting time increasing linearly with trial count (Figure 3F). We also simulated datasets with varying amounts of noise in the drift, with standard deviations ranging from 0 to 0.2. The model accurately recovered the systematic history bias with good precision as long as the noise in random drift was lower than sensory noise ($\sigma_d \le 1/\beta_s$, grey box). Beyond this regime, even though the model correctly fits observed choices (not shown) and correctly infers the direction of systematic updates, its estimates are biased and variable due to the high degree of indeterminacy in the data.}

\subsection{Comparing the model-free correction and the model-based approach}

\afterpage{
\refstepcounter{figure}
\begin{center}
\includegraphics[width=12.5cm]{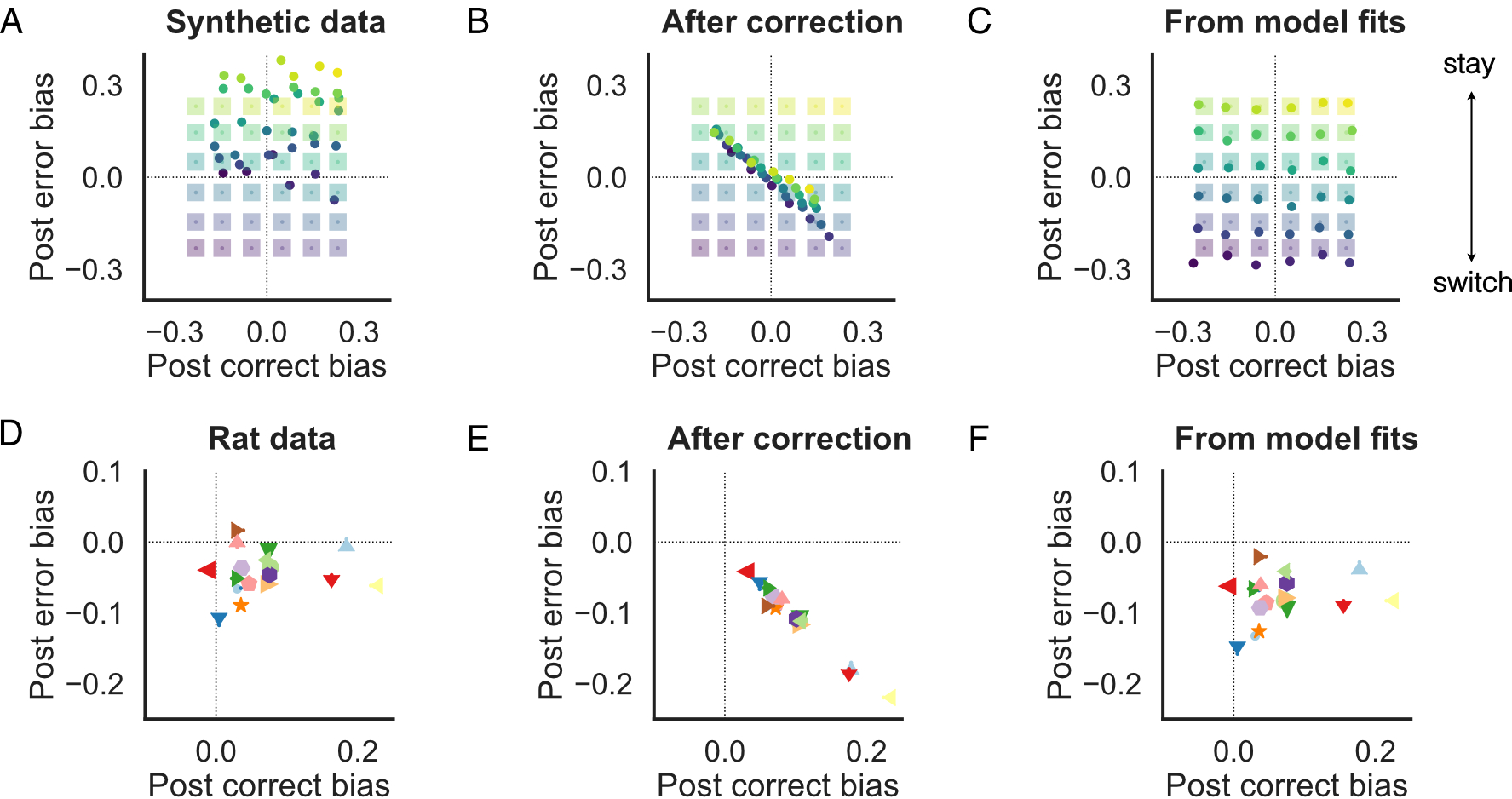}
\end{center}
\figcaption*{
\justifying
\textbf{FIGURE 4 : \add[DG]{The model-based approach but not MF correction preserves variability in history updating in synthetic and real datasets}}
\small{
\textbf{A. } \change[DG]{Biases}{History biases} \remove[DG]{measured by the difference between conditioned psychometric curves} following left and right wins (x-axis) or losses (y-axis) at the point of indifference \remove[DG]{(stimulus strength at which probability of choosing either option is equal)}. Squares denote the biases caused by systematic trial by trial updates, and dots denote the observed biases when slow drifts are present. \change[DG]{Colors represent unique parameter settings.}{Each color represents a unique setting of $\beta_C$ and $\beta_E$ parameters, which span a spectrum of stay (positive values) and switch (negative values) biases.}
\textbf{B.} Distribution of biases after applying MF correction 
\textbf{C.} Distribution of inferred history biases from fits to a model that includes slow drift and trial-by-trial updating. 
\textbf{D.} Distribution of observed history biases from 19 rats (dataset from \cite{brunton_rats_2013}). Symbols denote individual rats.
\textbf{E.} Distribution of biases after applying correction from \cite{lak_reinforcement_2020}. MF correction distorts the distribution, uniformly producing equal degrees of win-stay and lose-switch biases.
\textbf{F.} Distribution of inferred history biases from fits to a model that includes slow drift and trial-by-trial updating. The distribution is shifted relative to A), but individual rats retain their relative position and asymmetries between post-correct and post-error updating.
}}
\label{fig4}
\vspace{1cm}
}

Next, we compare the history biases estimated by the two approaches with synthetic as well as real data. We simulate choices for a variety of possible post-correct and post-error biases (\hyperref[fig4]{Figure 4A-C}, squares, see \hyperref[Table 1]{Table 1} in Methods) in \change[DG]{addition to}{presence of} random drifts in decision criteria and evaluate the results obtained by applying the MF correction and by model-fitting. \change[DG]{We summarize these history effects by plotting the choice bias observed post-correct trials against post-error trials at the point of indifference}{We summarize these history effects by plotting the choice history bias observed following correct trials against that observed following error trials, measured at the point of indifference i.e. the stimulus strength at which the fraction of rightward choices is 0.5}. First, we show that consistent with our previous simulations \change[DG]{-}{, the} presence of slow drifts exaggerates stay biases in the data (\hyperref[fig4]{Figure 4A}, dots) especially following error trials, and obscures the true underlying update strategy (\hyperref[fig4]{Figure 4A}, squares). Application of the MF correction does not selectively remove the influence of slow drifts, rather it warps all the history effects to lie along the diagonal represented by win-switch lose-stay and win-stay lose-switch strategies (\hyperref[fig4]{Figure 4B}, dots). In contrast, the model-based inference of systematic updates successfully removes the influence of slow drifts and reveals the underlying strategy for all considered settings (\hyperref[fig4]{Figure 4C}, dots vs squares). 

Next, we analyze choice data from 19 rats participating in a two-alternative auditory evidence-accumulation task (data from \cite{brunton_rats_2013}; see Methods for more details). In this study, even though successive trials were independent of each other, the rat subjects showed varying extents of win-stay and lose-switch biases (\hyperref[fig4]{Figure 4D}). The application of MF correction to this dataset, squashes the variability in history biases across all rats and returns a homogenous win-stay lose-switch bias (\hyperref[fig4]{Figure 4E}). \change[DG]{Whereas}{In contrast}, the model fits retain the individual variability (\hyperref[fig4]{Figure 4F}) \change[DG]{while alleviating the stay effects in post-error bias that are caused by slow drifts}{ and infer a higher post error switch bias compared to that observed in raw data by accounting for stay tendency that is induced by slow drifts} (\hyperref[fig4]{Figure 4A}).

\section{Discussion}
Nonstationarities in decision-relevant variables, if overlooked, can bias one’s estimates of psychophysical parameters and obscure strategies that underlie behavior \cite{frund_inference_2011}. Previous work \cite{lak_reinforcement_2020} has shown that unaccounted for slow drifts in decision criterion could obscure trial-by-trial updates produced by active learning, and proposed a correction to remove its influence (\hyperref[fig1]{Figure 1}).  Here we investigate the performance of this correction under a range of different generative models, and demonstrate that it fails to selectively remove the influence of slow drifts in the presence of systematic trial-by-trial updates of decision variables (\hyperref[fig2]{Figure 2}). Moreover, applying the correction corrupts the estimates of trial-history influences, biasing them towards a small subset of possible strategies. To mitigate these shortcomings, we propose an alternate approach of explicitly modeling slow fluctuations in the decision-making process (\hyperref[fig3]{Figure 3}). We demonstrate that this model-based approach can successfully disambiguate systematic updates from non-specific drifts, hence preserving the structure of individual variability in behavioral strategies (\hyperref[fig4]{Figure 4}).

Non-specific drifts in decision variables may arise from fluctuations in internal states such as attention, arousal and motivation, or from other internal sources of noise \cite{renart_variability_2014}. Hence, some authors have used neural measurements to gauge their dynamics \cite{ridderinkhof_errors_2003, eichele_prediction_2008, cowley_slow_2020}. Of particular interest is the work of Cowley and colleagues who leverage slow drifts in stimulus encoding in visual and prefrontal cortex to explain fluctuations in behavior and pinpoint the affected decision-variable. 

Even in the absence of such detailed measurements, it may be possible to account for unsystematic influences by modelling the noisy dynamics of unobserved variables, as we have done in this manuscript for the decision criterion (see \cite{smith_dynamic_2004} for formative work). This approach has also been used to identify systematic and random contributions to action value learning \cite{findling_computational_2019}, allowing for the decomposition of errors into noise-driven and choice-driven components.  

In general, it is likely that multiple decision variables drift over time, or that systematic updates occur over longer timescales, in which case it is important to incorporate those assumptions into the model. While the model we suggest can be readily extended to estimate additional forms of dynamics, other authors have put forward alternate approaches. Previous studies have proposed regression models to infer the trajectory of psychometric parameters over trials, regularizing their estimates with empirically determined priors \cite{bak_adaptive_2016, kattner_trial-dependent_2017, bak_adaptive_2018, roy_extracting_2021}. In another study authors consider a generative model in which the latent state which governs the setting of psychometric parameters undergoes discrete change \cite{ashwood_mice_2022}. In parallel, some generative models might allow for the use of filtering methods to recover signal from the observations. Indeed, previous studies have shown that low-pass filtering the sequence of choices using a moving average filter can adequately help estimate the slow drifts in bias \cite{cowley_slow_2020, mochol_prefrontal_2021} without inducing the kind of biases induced by the MF correction. The relative utility of these different approaches (state space models, regression, filtering, corrections) would depend on the desired level of explanatory power, number of data samples and efficiency of the inference algorithms among other factors. In any case, studies would benefit from explicit comparison between different hypothesized generative models. 
 
Outside the realm of trial-based behavior many exciting advances have been made in inferring the dynamics underlying naturalistic behaviour \cite{wiltschko_mapping_2015, eyjolfsdottir_learning_2016, sharma_point_2018, calhoun_unsupervised_2019}. In the future, it would be interesting to bring these advances into the study of learnt perceptual behaviour and develop inference strategies for more sophisticated generative models. Exploring the origins and logic of slow fluctuations inferred with such models might help advance our understanding of the principles that underlie behaviour and learning.


\section{Methods}
\subsection{Simulation Details}
We simulated a signal detection theory observer that compares a noisy sample of the stimulus (range 0 to 1, corrupted by logistic noise) to a decision criterion, making a rightward choice if the sample exceeds the criterion and leftward choice otherwise. 

The decision criterion $b_t$ of this observer varies from trial to trial relative to a fixed baseline $\beta_0$, due to unsystematic variations arising from random drift $z_t$ and systematic variations based on the choice and outcome of the previous trial $c_{t-1}, o_{t-1}$. The drift $z$ in decision criterion $ b $ over trials $t$ was simulated according to the following autoregressive process (discrete time Ornstein-Uhlenbeck process):
$$ z_t =  (1-\lambda) z_{t-1} + \sigma_d \epsilon_t $$
where $\epsilon_t$ is an i.i.d sequence of standard Gaussian random variables. Throughout the study the decay rate was fixed to a small value ($\lambda = 5e-4$) to prevent the criterion from drifting too far from the true boundary and the standard deviation of the Gaussian noise, $\sigma_d$ was set at $0.05$ when applicable. The systematic variations based on previous choice and outcome took the form of an additive bias to the decision criterion. The bias was $\beta_C/-\beta_C$ following rightward/leftward choices and positive outcomes, and $\beta_E/-\beta_E$ following rightward /leftward choices and negative outcomes. 
Therefore, the decision criterion on any trial was given by the following equation:
\begin{align*} 
b_t(c_{t-1}, o_{t-1}, z_t) = \beta_0 &+ z_t \\
&+ \beta_C \mathbb{I}(c_{t-1}=1, o_{t-1}=1) - \beta_C \mathbb{I}(c_{t-1}=0, o_{t-1}=1) \\
&+ \beta_E \mathbb{I}(c_{t-1}=1, o_{t-1}=0) - \beta_E \mathbb{I}(c_{t-1}=0, o_{t-1}=0) 
\end{align*} 
where $\mathbb{I}(c_{t-1},o_{t-1})$ denotes an indicator variable for a particular combination of previous choice $c_{t-1}$ and outcome $o_{t-1}$. The baseline value $\beta_0$  was set to produce an equal proportion of leftward and rightward choices when stimulus carried no information i.e. stimulus strength of 0.5. 

Finally, choice $c$ on trial $t$ depended upon the stimulus strength on the current trial $s_t$, and the decision criterion $b_t$. The probability of making a rightward choice was given by the logistic function:
$$ p(c_t = 1| s_t,b_t) = \frac{1}{1+e^{-(s_t - b_t)/\beta_S}}$$
\change[DG]{We simulated 40000 trials at a time, and the plotted}{The} psychometric functions were fit using the logistic regression function from the scikit-learn module in Python \cite{pedregosa_scikit-learn_2011}.

\vspace{5mm}

\begin{minipage}{\linewidth}
\refstepcounter{table}
\tabcaption*{TABLE 1 : Parameters used in simulations}%
\begin{tabular}{lcccccc}
\hline
&Number of trials &$\beta_S$ &$b_t$ &$\beta_C$ &$\beta_E$ &$\sigma_d$\\
\hline
Fig 1C-D&40,000&10&-5&0&0&0.05\\
Fig 1E-F&80,000&10&-5&0&0&0.10\\
Fig 2B-E solid (left to right)&80,000&10&-5&1,1,1,-1&-1,0,1,-1&0.00\\
Fig 2B-E dotted (left to right)&80,000&10&-5&1,1,1,-1&-1,0,1,-1&0.05\\
Fig 3B-D &40,000&10&-5&1&1&0.05\\
Fig 3E-F &see x-axis&10&-5&1&1&0.05\\
Fig 3G &20,000&10&-5&1&1&see x-axis\\
Fig 4A-C &40,000&10&-5&$\pm1, \pm0.6, \pm0.2$ &$\pm1, \pm0.6, \pm0.2$ &0.05\\
\hline
\label{Table 1}%
\end{tabular}
\end{minipage}

\subsection{History Bias and Correction}
History bias post-correct/error at each stimulus strength is computed as the difference between the probability of going right following rightward and leftward correct/error trials. Therefore, the history bias following correct trials is given by $$bias^{PC}_{t|t-1} = \frac{1}{2} \left (p(c_t = 1|s_t, o_{t-1} = 1, c_{t-1} = 1) - p(c_t = 1|s_t, o_{t-1} = 1, c_{t-1} = 0) \right) $$ And similarly, following error trials history bias is computed as $$\frac{1}{2} \left (p(c_t = 1|s_t, o_{t-1} = 0, c_{t-1} = 1) - p(c_t = 1|s_t, o_{t-1} = 0, c_{t-1} = 0) \right )$$

To correct for the influence of slow drifts, we subtract the acausal bias due to the past trial on the choice preceding it from the bias due to the past trial on current choice $bias_{t-2|t-1}$ \cite{lak_reinforcement_2020, mendonca_impact_2020}, i.e. following a rewarded trial the corrected bias for each stimulus strength was computed according to: 
\begin{align*}
bias^{PC}_{t|t-1} - bias^{PC}_{t-2|t-1} =  &\frac{1}{2}\big(p(c_t = 1|s_t, o_{t-1} = 1, c_{t-1} = 1) - p(c_t = 1|s_t, o_{t-1} = 1, c_{t-1} = 0)\big) - \\
&\frac{1}{2}\big(p(c_{t-2} = 1|s_{t=2}, o_{t-1} = 1, c_{t-1} = 1) - p(c_{t-2} = 1|s_{t-2}, o_{t-1} = 1, c_{t-1} = 0 \big) 
\end{align*} 
The corrected bias following error trials was similarly computed, but by conditioning on previous error trials i.e. $o_{t-1} = 0$.

\subsection{Model-based fitting}
We denote the choice on trial $t \in {1...T} $ as $c_t$ such that $c_t = 1$ if the choice is towards right and $0$ otherwise. The choice on any given trial is formed by comparing the value of a noisy perceptual sample (centered around the true stimulus $s_t$ with logistic noise) to the category boundary or criterion $b_t$, which we assume is time-varying and is modulated by trial-history. If the value of the sample exceeds the criterion, a rightward choice is made. Therefore, conditioned on the drifting criterion and the stimulus the choice of an agent on any given trial is given by a Bernoulli distribution with mean
$\mathbb{E}[c_t|b_t, s_t] = \frac{1}{1+ \text{exp}(-(\beta_S s_t - b_t))}$
where $\beta_S$ determines the sensitivity of an agent to the perceptual stimulus $s_t$.

We assume that the drift $z_t$ in criterion $b_t$ evolves according to a random walk with step size $\sigma_d$, we assume that the mean and variance of the initial state are known $(z_0 = 0, \sigma_0^2 = 0.01)$: 
\begin{align*}
z_1\sim \mathcal{N}(z_0, \sigma^2_0) \\
z_{t+1} | z_{t} \sim \mathcal{N}(z_t, \sigma_d^2)
\end{align*}
The decision criterion $b_t$ is the summation of a fixed baseline $\beta_0$, the drift, and trial-history influences:
\begin{align*}
b_t = z_t + \beta_0 + \beta_C(\mathbb{I}_{t-1}^{RC} - \mathbb{I}_{t-1}^{LC}) + \beta_E(\mathbb{I}_{t-1}^{RE}-\mathbb{I}_{t-1}^{LE})
\end{align*}
where $\mathbb{I}^{RC}_{t-1}$, $\mathbb{I}^{LC}_{t-1}$, $\mathbb{I}^{RE}_{t-1}$ and $\mathbb{I}^{LE}_{t-1}$ are indicator variables denoting successes or failures on the previous trial ($t-1$)  following rightward and leftward choices. $\beta_C$ and $\beta_E$ determine how much previous trial’s choice and outcome modulate the criterion.

This model for choice behavior is essentially a linear dynamical system with Bernoulli emissions. Therefore, to infer the parameters $\Theta = \{\sigma_d, b_0, \beta_S, \beta_C, \beta_E\}$ of this latent variable model we use an expectation-maximization  algorithm, similar to those described before (Smith and Brown 2003, Macke et al. 2011, Macke et al. 2015). For the E-step, we require the log of posterior distribution $P(\mathbf{z}|\mathbf{c}, \Theta)$ over the latent drift given the observed pattern of choices and our current estimate of the parameters $\Theta$:
\begin{align*}
\text{log}P(\mathbf{z}|\mathbf{c}, \Theta) = &\sum_{t=1}^T \left(c_t h_t  + \text{log}\left(1 - \frac{1}{1+ e^{-h_t}}\right) \right) - \frac{(z_1-z_0)^2}{2\sigma_0^2} - \frac{1}{2\sigma_d^2}\sum_{t=1}^{T-1}(z_{t+1}-z_t)^2 + \text{const}
\end{align*} 
where $h_t = \beta_S s_t -b_0 - z_t -\beta_C\mathbb{I}^C_t - \beta_E \mathbb{I}^E_t$. Owing to Bernoulli observations this distribution is not available in closed-form but is concave, so we approximate it with a multivariate Gaussian distribution (Laplace approximation) $P(\mathbf{z}|\mathbf{c}, \Theta) \sim \mathbb{N}(\mathbf{\mu^*}, \Sigma^*)$ where $\mathbf{\mu}*$ is a vector of size $T$ and is set to maximize the posterior $\mathbf{\mu}^* = \arg\max_{\mathbf{z}}P(\mathbf{z}|\mathbf{c}, \Theta)$ and $\Sigma^*$ is of size $T\times T$ and is set to the inverse Hessian of the log posterior evaluated at $\mathbf{\mu}^*$ i.e. $\Sigma^* = -[ \nabla^2_{\mathbf{z}} \text{log}P(\mathbf{z}|\mathbf{c}, \Theta) |_{\mathbf{z} = \mathbf{\mu^*}} ]^{-1}$.

The M-step updates parameters by optimizing the expected total data log-likelihood with respect to the parameters 
\begin{align*}
\Theta^* = \arg \max _{\theta} \int [\text{log}(\mathbf{c}|\mathbf{z}, \Theta) + \text{log}P(\mathbf{z}|\Theta)] \mathbb{N}(\mathbf{z}|\mathbf{\mu}^*, \Sigma^*) d\mathbf{z} 
\end{align*} 
We update the parameters using L-BFGS except $\sigma_d$ which we update analytically. We use the \textit{ssm package} \cite{linderman_ssm_2020} in Python for this inference.

\subsection{Dataset studied}
We analyze the trial-history effects of nineteen rats (50223 $\pm$ 21915 trials per rat) performing an evidence accumulation task, originally published in \cite{brunton_rats_2013}. In this task, the subjects were presented with two simultaneous streams of randomly-timed discrete pulses of evidence (clicks), one from a speaker to their left and the other to their right, for a predetermined duration. The subjects were required to maintain fixation throughout the entire stimulus, failure to do so led to a violation trial. At the end of the stimulus, the subjects had to orient towards the side which played the greater number of clicks to obtain a water reward. The discrimination difficulty was controlled on each trial by varying the ratio of the generative Poisson rates of the two click streams. Successive trial’s difficulty and rewarded side were independently sampled. Since the animals neither made a choice nor received an outcome on violation trials, we ignore them while computing trial-history effects. 

\subsection{Data and Code availability}
The code and data associated with this manuscript is available here: \url{https://doi.org/10.24433/CO.8821874.v1} as a standalone executable.

\section*{acknowledgements}
We thank Armin Lak, Tim Kim, Thomas Luo and Adrian Bondy for helpful discussion and comments on this manuscript.


\bibliography{sample}

\end{document}